\pdfoutput=1

\documentclass[aps,pra,twocolumn,superscriptaddress,floatfix,altaffilletter]{revtex4-1}

\usepackage[utf8]{inputenc}
\usepackage{amsmath,amsfonts,amssymb,amsthm}
\usepackage{graphicx}
\usepackage{subcaption}

\usepackage{xcolor}

\usepackage{siunitx}
\usepackage{tikz}
\usetikzlibrary{shapes, decorations.pathreplacing, positioning}
\usepackage{tikz-3dplot}

\def\mathclap#1{\text{\hbox to 0pt{\hss$\mathsurround=0pt#1$\hss}}}
\newcommand{\im}{\mathrm{i}}
\renewcommand{\vec}[1]{\mathbf{#1}}
\newcommand{\pvec}[1]{\mathbf{#1}_\parallel}
\newcommand{\vecUnit}[1]{\mathbf{\hat{#1}}}

\newcommand{\etal}{\textit{et~al.}}

\begin{document}


\title{Nanometer resolution mask lithography with matter waves: Near-field binary holography}



\author{Torstein Nesse}
\affiliation{Department of Physics, NTNU -- Norwegian University of Science and Technology, NO-7491 Trondheim, Norway}

\author{Ingve Simonsen}
\email{ingve.simonsen@ntnu.no}
\affiliation{Department of Physics, NTNU -- Norwegian University of Science and Technology, NO-7491 Trondheim, Norway}
\affiliation{Surface du Verre et Interfaces, UMR 125 CNRS/Saint-Gobain, F-93303 Aubervilliers, France}

\author{Bodil Holst}
\affiliation{Department of Physics and Technology, University of Bergen, All\'{e}gaten 55, 5007 Bergen, Norway}



\date{\today}


\begin{abstract}
Mask-based pattern generation is a crucial step in microchip production. The next-generation extreme-ultraviolet- (EUV) lithography instruments with a wavelength of \SI{13.5}{\nano\meter} is currently under development. In principle, this should allow patterning down to a resolution of a few nanometers in a single exposure. However, there are many technical challenges, including those due to the very high energy of the photons. Lithography with metastable atoms has been suggested as a cost-effective, less-complex alternative to EUV lithography. The great advantage of atom lithography is that the kinetic energy of an atom is much smaller than that of a photon for a given wavelength.
However, up till now no method has been available for making masks for atom lithography that can produce arbitrary, high resolution patterns. Here we present a solution to this problem. First, traditional binary holography is extended to near-field binary holography, based on Fresnel diffraction. By this technique, we demonstrate that it is possible to make masks that can generate arbitrary patterns in a plane in the near field (from the mask) with a resolution down to the nanometer range using a state of the art metastable helium source. We compare the flux of this source to that of an established EUV source (ASML, NXE:3100) and show that patterns can potentially be produced at comparable speeds. Finally, we present an extension of the grid-based holography method for a grid of hexagonally shaped subcells. Our method can be used with any beam that can be modeled as a scalar wave, including other matter-wave beams such as helium ions, electrons or acoustic waves.
\end{abstract}

\maketitle


\section{Introduction}
\label{sec:introduction}
In standard photolithography, the resolution is determined by the wavelength of the light: the smaller the wavelength, the higher the resolution. The present industrial photolithography standard is the immersion scanner using a \SI{193}{\nano\meter} light source. Following standard diffraction theory (Abbe resolution criterion) this light source gives a maximum resolution in air of \SI{95}{\nano\meter}. This is increased by use of off-axis illumination and a highly refractive immersion medium~\cite{ITRS}. Furthermore, in modern chip production the patterns are generated by subtle use of underexposure, overdevelopment, and multiple exposures so that patterns with a resolution of around \SI{20}{\nano\meter} can be created. Lithography methods for higher resolution exist; for example, electron-beam lithography, which is used to make the masks for photolithography. However, these are all serial lithography techniques and are much slower than mask-based lithography. The industry is currently implementing the next generation of lithography devices, and extreme-ultraviolet (EUV) lithography based on a \SI{13.5}{\nano\meter}-wavelength light source is expected to be able to produce patterns with a resolution of less than \SI{10}{\nano\meter} in single exposures~\cite{ITRS}.

Atom lithography has been suggested as an alternative to EUV lithography. For a given wavelength the energy of the atom is much less than the energy of the photon.
For instance, the energy of an EUV photon of wavelength $\lambda=\SI{13.5}{nm}$ is $E=hc/\lambda \approx \SI{91.8}{\electronvolt}$, where $h$ denotes Planck's constant and $c$ is the speed of light in a vacuum. On the other hand, a helium atom of the same wavelength ($\lambda=\SI{13.5}{nm}$) has a kinetic energy $E=h^2/(2m \lambda^2) \approx \SI{0.011}{meV}$, where $m$ is the mass of the helium atom. High-intensity atom beams with narrow velocity distributions can be created by expansion from a high-pressure reservoir through a nozzle followed by selection of the central beam with a conically shaped aperture, which prevents backstreaming into the beam. This aperture is typically referred to as the ``skimmer''~\cite{Pauly2000}. For helium atoms with kinetic energies between \SI{0.02}{\milli\electronvolt} (corresponding to a liquid-nitrogen-cooled beam) and \SI{0.06}{\milli\electronvolt} (room-temperature beam), the corresponding wavelengths are between \num{0.1} and \SI{0.05}{\nano\meter}. This makes atom beams, in principle, a very attractive candidate for high-resolution pattern generation. One approach in atom lithography is to use a beam of metastable atoms for the pattern generation. When a metastable atom hits the substrate, it decays, and the energy of the metastable state is transferred to the substrate \cite{Berggren95,Baldwin2005,Ueberholz2002}. In their seminal paper from 1995, Berggren~\etal~\cite{Berggren95} demonstrated pattern generation in a thiol-based resist using a beam of metastable argon atoms manipulated by a light-field mask. Since then numerous groups have experimented with atom lithography using either metastable noble-gas atoms and patterning in resist or direct deposition of atoms on substrates. The energy released when a metastable atom decays is about \SI{10}{eV} for argon and \SI{20}{eV} for helium~\cite{Colin1954,Baldwin2005}.

In most of these experiments the atomic beams were manipulated either by light or electrostatic fields~\cite{Berggren95,Gardner2017,Adams1994,Hinderthur1998}. The reason for this is that atoms at low energies, as they typically are in these beams, do not penetrate any substrates. Furthermore the metastable atoms decay when they impinge on a surface. Therefore it is not possible to use masks made on substrates as is done in photolithography. The pattern has to be generated by open areas in the substrate where the beam can go through, so any closed path (\textit{i.e.} a circle) would lead to the segment of the mask bounded by the closed path falling out. This limited the patterns that could be made in a mask configuration to essentially stripes and dots. Experiments have also been done that involve focusing of atom beams with lenses~\cite{Eder2017,Koch2008,Doak1999,Carnal1991,Eder15,Eder12,Patton2006,Barr2016,Reisinger2008}. This can be used for serial writing of arbitrary patterns. However, for mass-scale production serial writing is not a suitable method.

In 1996 Fujita \etal~\cite{Fujita1996} used a different approach. Instead of using light or electrostatic fields, they made a solid mask consisting of a distribution of uniformly sized holes, etched in a silicon nitride membrane. The hole distribution was calculated with the theory of grid-based binary holography developed by Lohmann and Paris~\cite{Lohmann1967}, and later by Onoe and Kaneko~\cite{Onoe1979}: grid-based binary holography imposes the limitation on the binary holograms that the openings are all of the same size and positioned at specific positions of a rectangular grid. That is to say, the holes are not only uniformly sized, but are also placed at a regular minimum spacing. The hole distribution is an approximated Fourier transform of the final, desired pattern. Murphy and Gallagher~\cite{Murphy1982} extended the binary-holography technique of Lohmann and Paris~\cite{Lohmann1967} to work also for hexagonal grids. Originally, the binary-holography method was developed to create holograms for electromagnetic waves with use of a computer, and the procedure is often referred to as ``computer-generated holography'' in the literature. Because of the de Broglie wavelength associated with a matter wave, the method also works for atom beams. It may be necessary to include a correction caused by the van der Waal interaction between the mask material and the atoms. However, as shown in a range of experiments~\cite{Grisenti1999} the only effect will be a slightly smaller effective hole size, which can easily be corrected for. Thus, we do not discuss this further in this paper.

Because the phase of the atoms when they arrive at the image plane (target plane) is not important, only the intensity is, many different hole distributions can create the same intensity pattern. In a recent publication it was shown how it is possible to vary the number of open holes in a mask over a large range without changing the final pattern~\cite{Nesse2017-2}. Coverage differences of up to \SI{83}{\percent} were demonstrated.

Until now one major problem with the binary-holography method has been that it is based on monochromatic, plane incident and outgoing waves. The binary mask is generated on the basis of a Fourier transform of the target pattern, which is assumed to be located at infinity. This means that the standard binary-holography method cannot be used to make patterns with high spatial resolution without the introduction of a lens that draws the Fraunhofer diffraction pattern in from infinity. Some work has been done on atomic lenses as mentioned earlier but no lenses with the required precision presently exist. Furthermore real atom sources are not perfectly plane waves and they are not perfectly monochromatic. The monochromaticity and spatial coherence of an atom beam are determined by the velocity distribution (wavelength distribution) and extension of the source~\cite{Patton2006}. The ultimate coherent beam would seem to be a Bose-Einstein condensate (BEC). Recently, Keller~\etal~\cite{Zeilinger14} generated a beam of BEC metastable helium atoms. However, standard Fraunhofer diffraction theory does not apply for a BEC \cite{Fouda2016}. Furthermore, the de Broglie wavelength of a dropping BEC of helium is very large, about \SI{30}{\nano\meter} after a drop of \SI{0.5}{\meter}. For high-resolution lithography, one wants to use short wavelengths, since the wavelength determines the ultimate target pattern resolution that can be achieved. One can think of experimental ways to get around this (e.g., by moving the mask relative to the BEC so that the BEC wavelength relative to the mask becomes smaller) but considerable amendments would have to be made to the theory we present here. It is also very challenging to make a high-flux BEC source. It should be mentioned, however, that considerable progress has been made in this field in recent years~\cite{Bolpasi2014,Bennetts2017}.

Recently, a beam of metastable helium atoms with a very narrow wavelength distribution $\lambda/\Delta\lambda = 200$ was produced with a pulsed source~\cite{Even2015}. In this paper we show that it is possible to make binary masks that can be used to create patterns with nanometer resolution in a target plane close to the mask with use of an atom source with the wavelength distribution given above and no lenses.

\medskip
The rest of this paper is organized as follows. Section~\ref{sec:theory} presents the theoretical background of the method that we use for the creation of the mask producing the desired pattern on the screen behind the mask. In particular, the method consists of an approximate backpropagation step based on Fresnel diffraction for plane waves~(Sec.~\ref{sec:fresnel}), the mask-generation step using binary holography on hexagonal (or rectangular) lattices~(Sec.~\ref{sec:gbh_and_hex}), and an evaluation step where our mask design is put to the test by application of rigorous forward-propagation techniques for the propagation of the field of realistic sources through the mask and onto the screen~(Sec.~\ref{sec:FK}). The results obtained by this approach are presented and discussed in Sec.~\ref{sec:Results}. Section~\ref{sec:Throughput} contains a discussion of the throughput of helium atoms that can be expected for realistic experimental parameters. In particular, we estimate the writing speed (throughput) that can be achieved with a state-of-the-art metastable-atom source and compare it with that of an established EUV source (NXE:3300, ASML). Finally, the conclusions that we draw from the study are presented in Sec.~\ref{sec:Conclusions}. The Appendix details the binary-holography technique applied together with hexagonal grids.


\section{Theory}
\label{sec:theory}

\begin{figure}[tbh]
  \centering
  \includegraphics[width=0.98\columnwidth]{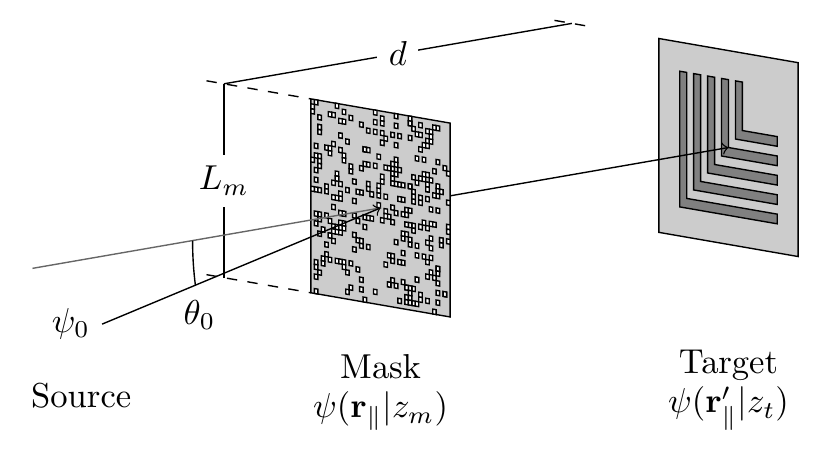}
  \caption{\label{fig:system} Overview of the lithography setup.
    A coordinate system is defined so that the positive $z$ direction is normal to the mask and target planes (which are parallel) and pointing away from the source (indicated by the central line). The source field, $\psi_0$,  is incident at an angle $\theta_0$ on the mask, and just after propagating through it, the mask field is $\psi(\pvec{r}|z_m)$, where $z=z_m$ defines the mask plane. The target field, $\psi(\pvec{r}'|z_t)$, is obtained after propagation of the mask field to the target plane ($z=z_t$) which is separated from the mask plane by distance $d=z_t-z_m$.
    The vectors $\pvec{r}$ and $\pvec{r}'$ are both perpendicular to the $z$ direction and therefore parallel to the mask and target planes. The pattern shown in the target plane is one of the standard test patterns used in lithography~\cite{Chang2001}.}
\end{figure}

The lithography system that we consider is presented in Fig.~\ref{fig:system}. It consists of a source, a mask that will diffract the incident wave propagating from the source, and a target plane (or screen) where the diffraction pattern will be displayed. In Fig.~\ref{fig:system}, these elements of the setup are labeled ``source'', ``mask'' and ``target'', respectively. It is assumed that the mask and the target planes are parallel. In lithography one aims at creating in the target plane, typically with high resolution, a predefined pattern --- called the ``target pattern'' in the following. We assume that such a target pattern has been identified and we stress that this pattern, in principle, can be arbitrary. However, in the target plane in Fig.~\ref{fig:system}, for the purpose of illustration, we use a structure consisting of a series of slightly displaced L shapes. This pattern is one of the standard resolution test patterns used in lithography~\cite{Chang2001}.

After the target pattern is chosen, the first step in the process of creating a binary holographic mask that corresponds to it consists in calculating the field just behind the mask plane (seen relative to the source). This field is referred to as the ``mask field'' in the following. It has the property that when this field is propagated to the target plane and the corresponding intensity is calculated, the target pattern is obtained (shown on the right in Fig.~\ref{fig:system}). To calculate the mask field, one therefore has to \emph{backpropagate} the field from the target plane to the mask plane. The field in the target plane --- the target field --- is taken as the square root of the intensity distribution of the target pattern multiplied by an arbitrarily chosen phase function that can vary over this plane. By changing the phase function, one obtains different masks that in principle should give rise to the same intensity distribution of the target pattern. The backpropagation of the target field can be performed in a number of different ways; which approach is the most appropriate depends on the structure of the target pattern and the separation between the mask and target planes (indicated as $d$ in Fig.~\ref{fig:system}). Here our goal is to create high-resolution masks (less than \SI{10}{\nano\meter}) that can be used in realistic lithographic setups based on neutral-atom beams. We therefore focus on geometries for which the source, mask, and screen (the target plane) are relatively close together;that is, the total (linear) system dimension is less than \SI{2}{\meter}. Under this assumption, the backpropagation can be performed with the use of (near-field) {\em Fresnel propagation}~\cite{Goodman2005}. For our purpose of generating high-resolution patterns, one cannot, for instance, rely on (far-field) Fraunhofer propagation~\cite{Goodman2005} which is used in existing binary-holography techniques. Therefore, to build a binary-holography technique around Fresnel propagation is novel and its use is prompted by the requirement of obtaining high resolution. The technique that is proposed here is called ``near-field binary holography'' in order to  distinguish it from ``standard'' binary holography, which assumes the target plane is in the far-field.

In the second step of the design process, the calculated mask field is used as the starting point for generation of the mask. The structure of the mask is designed so that when an incident field, originating at the source, passes through it, the field just behind the mask approximately equals the mask field calculated during the first (backpropagation) step.

The third and final step of the design process we propose involves evaluation of the performance of the generated mask to verify that it is capable of producing the desired target pattern. Numerical simulations are used for this purpose. Adequate results can be obtained with the use of the Fresnel-propagation approach similar to what is done in the first step to calculate the mask field. However, in contrast to what is done in the first step, we now forward propagate the incident field from the source, through the mask and onto the screen, where the corresponding intensity distribution is compared with the target pattern the design started from (see Fig.~\ref{fig:system}).

The subsequent discussion does not include the effect of gravity because gravity plays a minor role for the system that we study. To estimate the effect of gravity, let us start by assuming a typical velocity of the atoms at the mask plane of $v_0\sim \SI{E3}{m/s}$ and $d=\SI{50}{\micro\meter}$ for the mask-target separation (see Fig.~\ref{fig:system}). Now, if gravity acts parallel to the direction of flight, the velocity of the atoms will change during their flight. A straightforward calculation based on the kinematic equations for constant acceleration estimates the change in velocity during the flight from the mask to the target plane to be $\Delta v \approx gd/v_0$, where $g=\SI{9.8}{m/s\squared}$ is the acceleration due to gravity. With the numerical values assumed, we obtain $\Delta v \sim \SI{E-7}{m/s}$, which is $10$ orders of magnitude less than the velocity at the mask, and much less than the velocity spread of the source. 
On the other hand, when gravity acts perpendicular to the direction of flight, the beam will bend. The vertical displacement caused by gravity over distance $d$ is $g(d/v_0)^2/2 \approx \SI{1.25E-3}{nm}$, which is negligible compared with both the wavelength and the spatial extension of the source. Hence, we conclude that the effect of gravity can be ignored in our analysis.

\smallskip
In the subsequent subsections we detail each of the individual steps of the mask design and evaluation process.

\subsection{\label{sec:fresnel}Mask-field calculation: Fresnel propagation}
To calculate the mask field that we want our incident field to approximate after passing through the holographic mask, we need a way to propagate the desired target field backward from the screen (or target plane) to the mask. An accurate approximation for describing near-field propagation of scalar fields is the Fresnel diffraction integral~\cite{Goodman2005}. A coordinate system is defined so that the positive $\vecUnit{z}$ axis is pointing along the center of the illuminating beam (indicated by the central line in Fig.~\ref{fig:system}). Perpendicular to this direction the parallel mask and target planes are defined by $z=z_m$ and $z=z_t$, respectively, where $z_m$ and $z_t$ are arbitrary positive constants for which $z_m<z_t$~(Fig.~\ref{fig:system}). Let $\psi(\pvec{r}|z_m)$ denote the scalar field at the point $\vec{r}=\pvec{r}+z_m\vecUnit{z}$ of the mask plane. Here $\pvec{r}=(x,y,0)$ and a caret over a vector indicates that it is a unit vector. Similarly, $\psi( \pvec{r}'|z_t)$ represents the field at the point  $\vec{r}'=\pvec{r}'+z_t\vecUnit{z}$ of the target plane where $\pvec{r}'=(x',y',0)$. These two scalar fields, indicated in Fig.~\ref{fig:system}, are related by the Fresnel diffraction integral, which for the plane parallel geometry assumed here, takes the form~\cite{Goodman2005}
\begin{align}
  \label{eq:Fresnel-integral}
  \psi(\pvec{r}|z_m)
  &=
    \frac{e^{\im kd}}{\im\lambda d}
    e^{\im\frac{k}{2d} r_\parallel^2}
    \int 
    \mathrm{d}^2 r_\parallel'
    \left\{\psi( \pvec{r}'|z_t) e^{\im\frac{k}{2d}r_\parallel'^2} \right\}
    e^{-\im\frac{k}{d} \pvec{r}\cdot\pvec{r}' },
\end{align}
where $d=z_t-z_m$ denotes the distance between the mask and target planes as indicated in Fig.~\ref{fig:system}, $k=2\pi/\lambda$ is the wave number of the beam of wavelength $\lambda$, and the integration over $\pvec{r}'$ is assumed to extend over the entire target plane. Equation~\eqref{eq:Fresnel-integral} states that the scalar field in the mask plane, $\psi(\pvec{r}|z_m)$, can be obtained from the Fourier transform of the function $\psi( \pvec{r}'|z_t) \exp({\im\frac{k}{2d}r_\parallel'^2})$, which involves the target field $\psi( \pvec{r}'|z_t)$, by first evaluating the Fourier transform of this function for wave vector $\pvec{K}=k\pvec{r}/d$ and then multiplying the result by a known prefactor. The theoretical foundation of Fresnel propagation is based on Eq.~\eqref{eq:Fresnel-integral}.

When Fresnel propagation is performed, high numerical performance can be achieved due to the use of the fast Fourier transform. To this end, we first discretize the spatial coordinates of the mask and target planes. A flexible way of doing this was described by Muffoletto \etal~\cite{Muffoletto2007}. This method allows the calculation of Fresnel propagation between two areas on parallel planes that have the same number of discretization points, but can be scaled and shifted freely within the valid region of the Fresnel approximation. Therefore, one can use the method to propagate the field from the target plane to the mask plane as defined in Fig.~\ref{fig:system}.

The method starts by discretizing the in-plane spatial coordinates $\pvec{r}=(x,y,0)$ of the mask plane,
\begin{subequations}
  \label{eq:shifted_coordinates}
\begin{align}
  x_m &= x_0 + m\Delta x,
        \qquad
        0 \leq m \leq M-1
  \\
  y_n &= y_0 + n\Delta y,
        \;\qquad
        0 \leq n \leq N-1,
	\label{eq:shifted_coordinates-A}
\end{align}
and $\pvec{r}'=(x',y',0)$ of the target plane,
\begin{align}
  x'_p &= x'_0 + p\Delta x',
         \qquad
         0 \leq p \leq P-1
  \\
  y'_q &= y'_0 + q\Delta y',
         \qquad 
         0 \leq q \leq Q-1.
	\label{eq:shifted_coordinates-B}
\end{align}
\end{subequations}
Here $M$, $N$, $P$, and $Q$ are all positive integers; $x_0$, $y_0$ and $x_0'$, $y_0'$ are known offset parameters; and $\Delta x$, $\Delta y$ and $\Delta x'$, $\Delta y'$ are the discretization intervals in the mask and target planes, respectively. By substituting the results Eqs.~\eqref{eq:shifted_coordinates} into Eq.~\eqref{eq:Fresnel-integral} and  defining
\begin{align}
  U(m,n) &\equiv \psi(x_0 + m\Delta x, y_0 + n\Delta y | z)
  \\
  u(p,q) &\equiv \psi(x'_0 + p\Delta x', y'_0 + q\Delta y' | z' ),
\end{align}
one obtains the discretized version of the Fresnel diffraction integral~\eqref{eq:Fresnel-integral}, which we write in the form
\begin{widetext}
\begin{align}
  U(m,n)
  &=
    \frac{e^{\im kd}}{\im\lambda d}
    e^{\im\frac{k}{2d}\left(x_m^2 + y_n^2\right)}
    e^{-\im\frac{k}{d}\left(x_0'm\Delta x + y_0'n\Delta y\right)}
    \Delta x' \Delta y'
    \notag
  \\
  & \quad 
    \times 
    \sum\limits_{p=0}^{P-1}\sum\limits_{q=0}^{Q-1}
    \left\{
    u(p,q)
    e^{\im\frac{k}{2d}\left({x'_p}^2+{y'_q}^2\right)}
    e^{-\im\frac{k}{d}\left(x_p'x_0+y_q'y_0\right)}
    \right\}
    e^{-\im\frac{k}{d}\left(\Delta x'\Delta x\, pm + \Delta y'\Delta y\, qn\right)}.
    \label{eq:shifted_fresnel}
\end{align}
\end{widetext}
Here $U(m,n)$ and $u(p,q)$ are the discretized mask and target fields, respectively.

Equation~\eqref{eq:shifted_fresnel} has the structure of a \emph{scaled} two-dimensional discrete Fourier transform of the function inside the curly brackets, with the scaling parameters $s=(k/d)\Delta x'\Delta x $ and $t=(k/d)\Delta y'\Delta y$.  For a specific selection of scaling parameters, Eq.~\eqref{eq:shifted_fresnel} is in the form of the normal discrete Fourier transform, but we want to be able to select the two coordinate systems freely. Muffoletto \etal~\cite{Muffoletto2007} describe a technique for evaluating scaled discrete Fourier transforms of this kind by taking advantage of results due to Bailey and Swarztrauber~\cite{Bailey1991}. This technique is based on the rewriting of Eq.~\eqref{eq:shifted_fresnel} as a discrete convolution, which can be computed efficiently by performing three fast Fourier transforms.

The only major restriction coming from evaluating the Fresnel diffraction integral by the method of Muffoletto~\etal~\cite{Muffoletto2007} is the requirement that the number of discrete elements in the mask plane and in the target plane must be the same; that is, $M=P$ and $N=Q$. However, this limitation can be overcome at the cost of having to perform several Fresnel diffraction steps and shifting either the input region or the output region and tiling the results as described in Ref~\cite{Muffoletto2007}.

\subsection{\label{sec:gbh_and_hex}Mask generation: grid-based holography}

In the preceding subsection we outlined how to calculate the mask field that corresponds to a given target field. Here, given a mask field, we describe how a binary-holography mask can be constructed so that just after an incident beam passes through it, the resulting field will approximately equal the desired mask field.

Since this approach was recently presented in great detail in Ref.~\cite{Nesse2017-2}, here we will limit ourselves and give only the main steps that the method involves; for the necessary details and the mathematics, we refer the reader to Ref.~\cite{Nesse2017-2}. The method for generating masks that are able to transform the incident field, after passing through them, into the mask field, is based on the seminal work on binary holography of Lohmann and Paris~\cite{Lohmann1967} and Onoe and Kaneko~\cite{Onoe1979}. It starts by discretizing the mask field onto a rectangular grid of points that defines a set of nonoverlapping \textrm{cells} (or regions) filling a large portion of the mask plane. Next each of these cells is subdivided into an array of rectangular \textrm{subcells}. Some of these subcells are open so that the incident scalar field can be transmitted through them. Depending on which subcells are open and which are closed, the magnitude and phase of the field propagating away from the mask can take on a \textrm{finite} number of possible values. How many possible combinations there are depends on the size of the subcell array (see Ref.~\cite{Nesse2017-2}). Furthermore, which value the field will have depends on the configuration of open and closed subcells that the cell has; this is explained in great detail in Ref.~\cite{Nesse2017-2}. For instance, the total area of the open subcells determines the magnitude of the field associated with a cell.  On the other hand, the phase of the same field changes along only one of the subcell axes of the cell, so the positions of the openings along this direction allow one to modify the phase of the field propagating from the cell toward the screen in a plane spanned by this direction and the normal vector to the mask plane. So to construct a mask that is intended to form a given pattern in the target plane when a beam is transmitted through it and propagating away from it in a given direction, one has, for each cell, to choose the open-and-closed subcell configuration that corresponds to a field value that is the closest to the sampled-mask-field value for that cell. How such an optimization can be done in an efficient manner is described in  Ref.~\cite{Nesse2017-2}. The whole mask is designed by repetition of this process for each of its cells~\cite{Lohmann1967,Nesse2017-2}.

Murphy and Gallagher~\cite{Murphy1982} extended the binary-holography technique by placing the rectangular cells on a hexagonal grid.
This means that every other row of cells is shifted. In this paper we present results with cells on rectangular or  hexagonal grids, and we use both a rectangular grid and hexagonal grid of subcells. To facilitate the generation of masks based on a hexagonal grid of holes, we extend the grid-based holography method to work with a hexagonal grid of hexagonally shaped subcells. This extension is described in detail in the Appendix.

\subsection{\label{sec:FK}System evaluation: Huygens-Fresnel diffraction integral}

After constructing a realization of the mask, one can simulate the target pattern on the screen (the target plane) that the mask gives rise to by using the Fresnel diffraction integral, Eq.~\eqref{eq:Fresnel-integral}, as was done to find the mask field. Alternatively, a more rigorous approach based on the Huygens-Fresnel diffraction integral can be used~\cite{Goodman2005}. For the geometry that we consider, it is defined as 
\begin{align}
  \psi(\pvec{r}'|z_t)
  &=
    \frac{d}{\im\lambda}\int\limits_\Sigma \mathrm{d}^2 r_\parallel \,
    \frac{\exp(\im kR)}{R^2}\psi_{0}(\pvec{r}|z_m), 
	\label{eq:huygensfresnel}
\end{align}
where the integration domain $\Sigma$ is defined as the union of all holes of the mask. Moreover, the distance between the points $\vec{r}=\pvec{r}+z_m\vecUnit{z}$ and $\vec{r}'=\pvec{r}'+z_t\vecUnit{z}$ in the mask and target planes, respectively, is denoted
\begin{align}
	R &= \sqrt{(\pvec{r}' -\pvec{r} )^2 + d^2}.
\end{align}
In writing Eq.~\eqref{eq:huygensfresnel}, we have defined 
the incident field $\psi_{0}(\pvec{r}|z)$ that the source generates, and  evaluated it at $\vec{r}=\pvec{r}+z_m\vecUnit{z}$. An explicit expression for this field is given later [Eq.~\eqref{eq:source_description}].

The integral in Eq.~\eqref{eq:huygensfresnel} is evaluated numerically for each point of the target pattern by our integrating the expression over the area of each hole of the mask using an adaptive integration scheme. The integration is performed in such a way that a certain convergence criterion has been achieved.

In the simulations we model the supersonic helium source by using an adapted version of the \textit{virtual source model} introduced by Beijerinck and Verster~\cite{Beijerinck1981}. Recently, the same source model was used successfully to describe experimental measurements of the scattering of a beam of helium atoms from a photonic crystal structure~\cite{Nesse2017-1}. The virtual source model is based on the idea that after an initial region behind the nozzle where the atoms collide, they will eventually reach a free-flow regime at a distance from the nozzle referred to as the ``quitting surface.'' When this happens, the individual trajectories can be traced back to a plane that is perpendicular to the mean direction of travel and where the width of the spatial distribution function of the trajectories is a minimum~\cite{Beijerinck1981,DePonte2006}.

We now turn to presenting an explicit expression for $\psi_{0}(\pvec{r}|z)$. We consider the incident beam as an \textit{incoherent} and weighted \textit{superposition} of spherical waves or point sources, located approximately in the skimmer plane. The weight (or amplitude) used in the superposition is taken as a Gaussian function whose width, $\sigma$, mimics the half width of the skimmer. Mathematically the incident field at $\vec{r}=\pvec{r}+z\vecUnit{z}$ can be written as
\begin{align}
  \psi_{0}(\pvec{r}| z ) 
	&=
	\int\limits\mathrm{d}^2r^\star_\parallel \, 
	\frac{e^{-\frac{{r^\star_\|}^2}{2\sigma^2}}}{\sqrt{2\pi\sigma^2}}
	\frac{e^{\im k | \vec{r}-\vec{r}^\star |}}{|\vec{r}-\vec{r}^\star|}
	e^{\im \phi(\vec{r}^\star_\|)},
	\label{eq:source_description}
\end{align}
where $\vec{r}^\star_\|$ denotes a position in the skimmer plane, with the center of the skimmer opening at the origin. The integral in Eq.~\eqref{eq:source_description} should be evaluated over the entire skimmer plane. However, numerically we introduce a cutoff after the Gaussian factor becomes small (after a few standard deviations). In Eq.~\eqref{eq:source_description} $\phi(\vec{r}^\star_\|)$ represents a random phase function associated with the spherical wave source at $\vec{r}^\star_\|$. We assume that the random phase function is an uncorrelated stochastic variable that is uniformly distributed on the interval $[0,2\pi)$. The amplitude of the wave has been set to $1$ in Eq.~\eqref{eq:source_description}. To perform numerical simulations using the virtual source as an incident field, we must average the calculated diffraction patterns in the target plane (the screen)  over an ensemble of realizations of the random phase function.


\section{Results and discussion}
\label{sec:Results}

\begin{figure*}[htbp]
	\centering
    \begin{subfigure}[b]{0.4\textwidth}
		\includegraphics[width=\columnwidth]{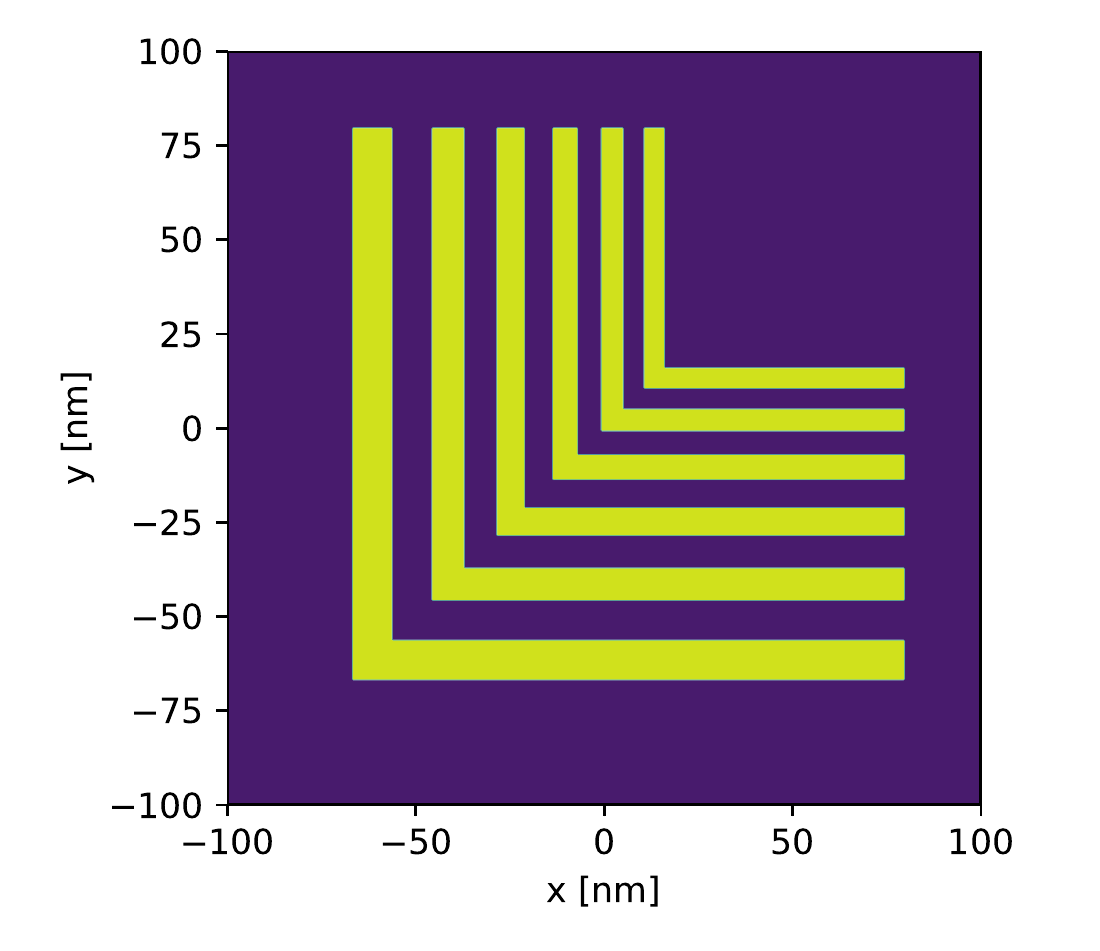}
        \caption{Corner lines}
    \end{subfigure}
    \begin{subfigure}[b]{0.4\textwidth}
		\includegraphics[width=\columnwidth]{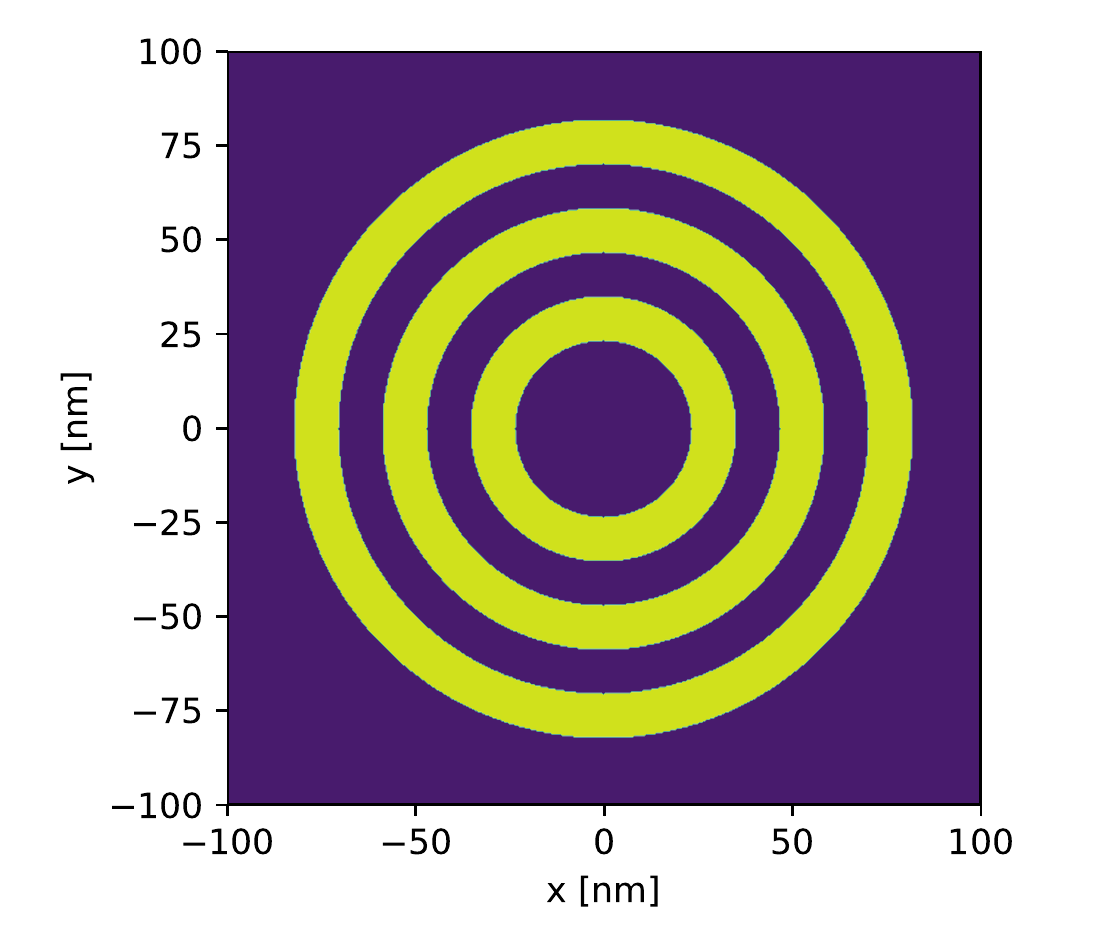}
        \caption{Circles}
    \end{subfigure}
	\caption{\label{fig:target}The target patterns. (a) Corner lines commonly used in lithography for resolution-test purposes~\cite{Chang2001}. The widths of the lines vary successively from \SI{10}{\nano\meter} to \SI{5}{\nano\meter}. (b) Concentric circles.}
\end{figure*}

\begin{figure*}[htbp]
    \centering
    \begin{subfigure}[b]{0.4\textwidth}
        \includegraphics[width=\textwidth]{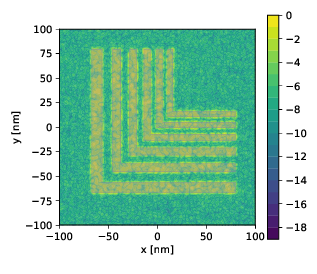}
        \caption{Intensity in target plane}
    \end{subfigure}
    \begin{subfigure}[b]{0.4\textwidth}
        \includegraphics[width=\textwidth]{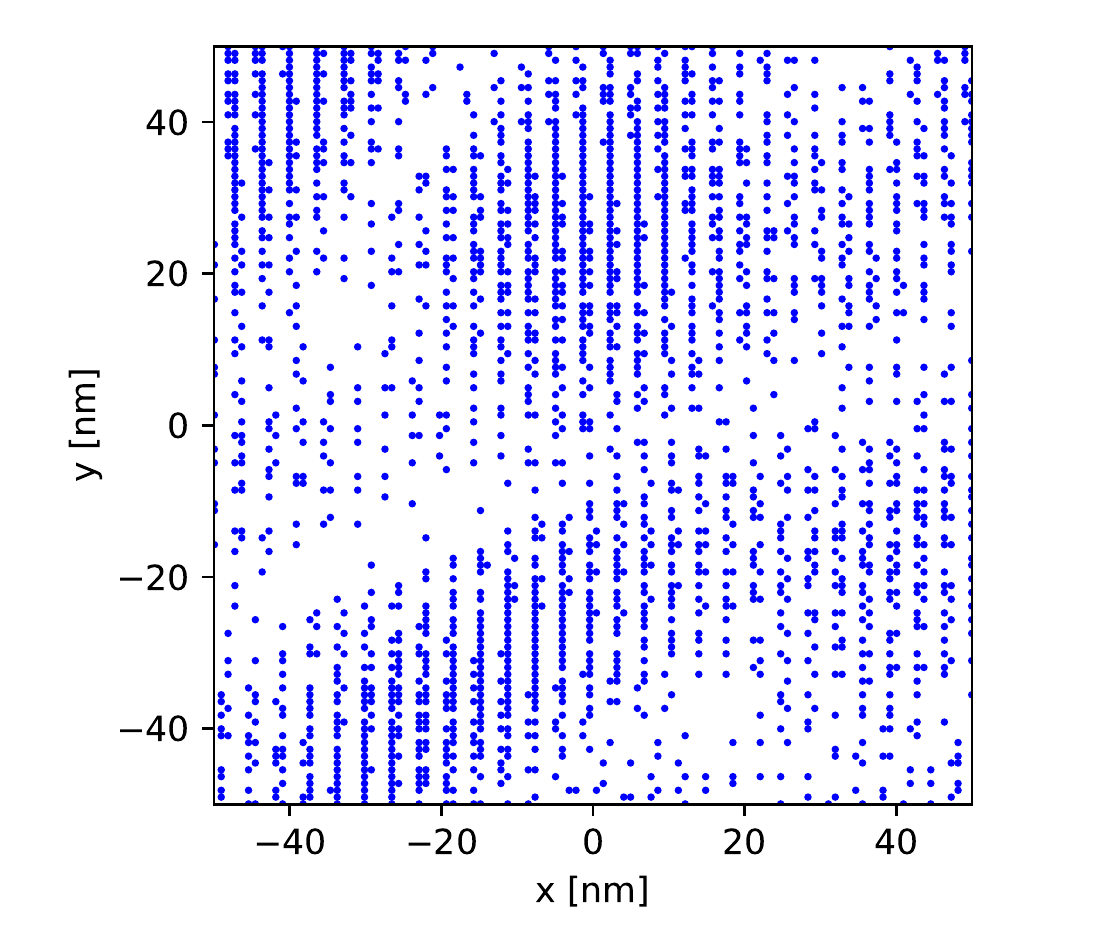}
        \caption{Mask cutout}
    \end{subfigure}
    \caption{\label{fig:kirchhoff_rect1} (a) Natural logarithm of the normalized intensity in the target plane obtained by simulations for a rectangular mask illuminated by a plane-wave source. The mask has a periodicity of \SI{0.9}{\nano\meter}, and the distance between the mask and target plane is \SI{40}{\micro\meter}. The desired target pattern is superimposed on the resulting target pattern, showing that the reproduction is true to size to within \SI{2}{\nano\meter}. (b) A $\num{100}\times\SI{100}{\nano\meter^2}$ cutout of the center of the mask.}
\end{figure*}

\begin{figure*}[htbp]
  \centering
  \begin{subfigure}[b]{0.4\textwidth} \includegraphics[width=\textwidth]{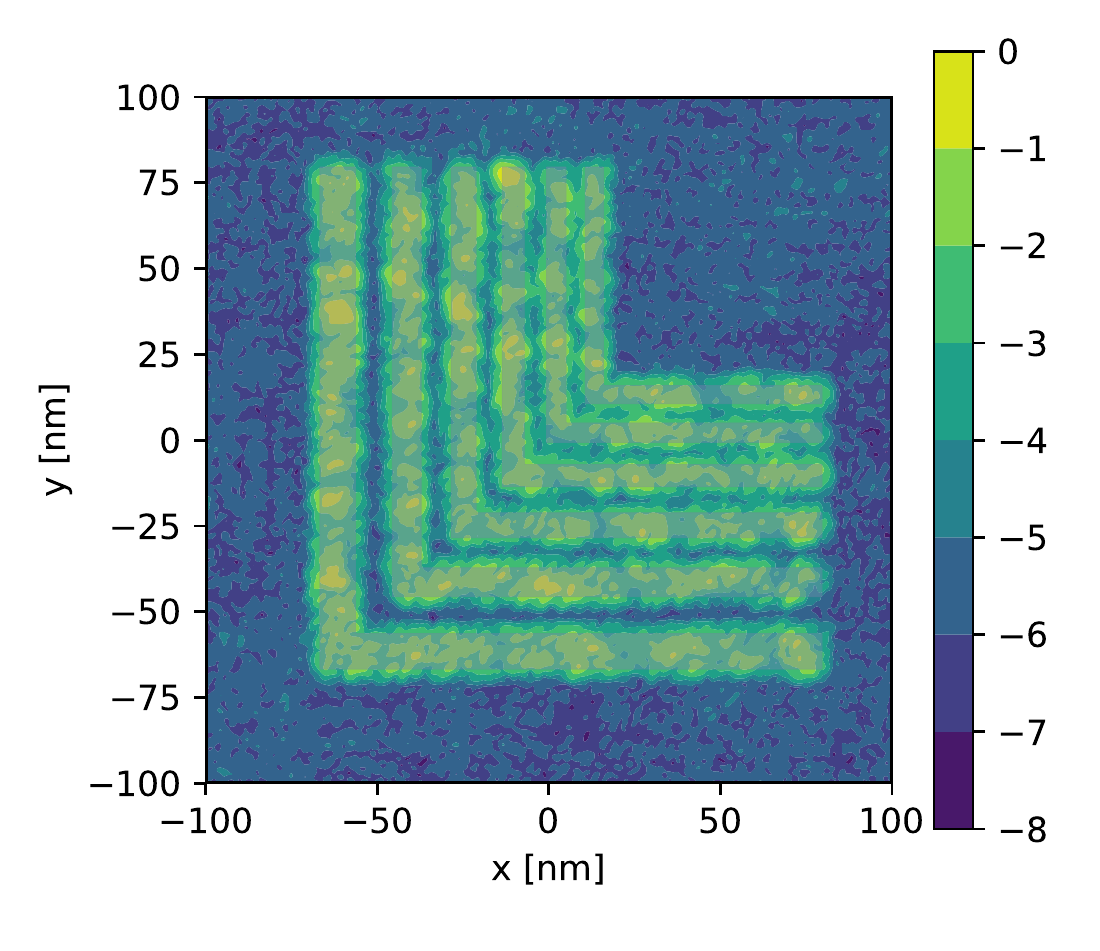}
    \caption{Intensity in target plane}
  \end{subfigure}
  \begin{subfigure}[b]{0.4\textwidth}  \includegraphics[width=\textwidth]{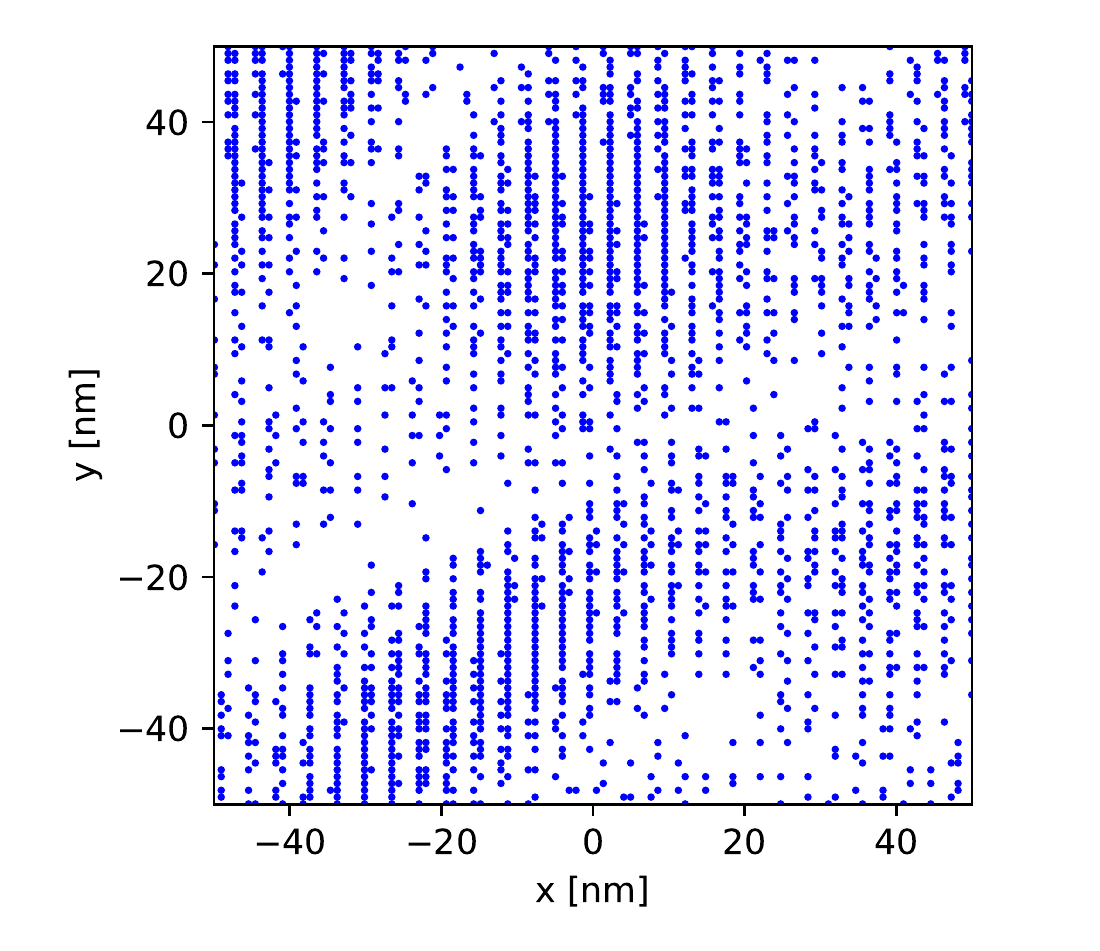}
    \caption{Mask cutout}
  \end{subfigure}
  \caption{\label{fig:kirchhoff_rect3} (a) Natural logarithm of the normalized intensity in the target plane obtained by simulations for a rectangular mask illuminated by a realistic source. The mask has a periodicity of \SI{0.9}{\nano\meter}, and the distance between the mask and target plane is \SI{40}{\micro\meter}. The desired target pattern is superimposed on the resulting target pattern, showing that the reproduction is true to size to within \SI{3}{\nano\meter}. (b) A $\num{100}\times\SI{100}{\nano\meter^2}$ cutout of the center of the mask.}
\end{figure*}

\begin{figure*}[htbp]
  \centering
  \begin{subfigure}[b]{0.4\textwidth}	\includegraphics[width=\textwidth]{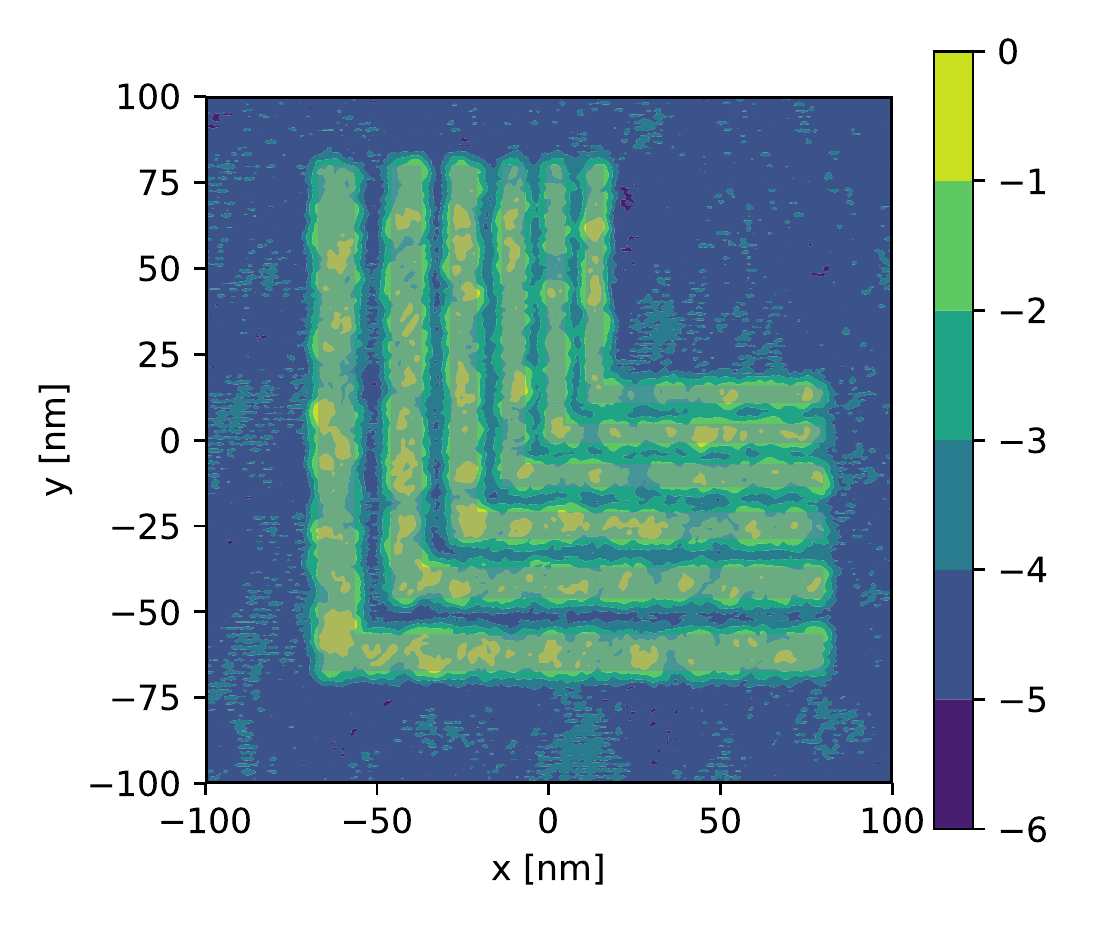}
    \caption{Intensity in target plane}
  \end{subfigure}
  \begin{subfigure}[b]{0.4\textwidth}	\includegraphics[width=\textwidth]{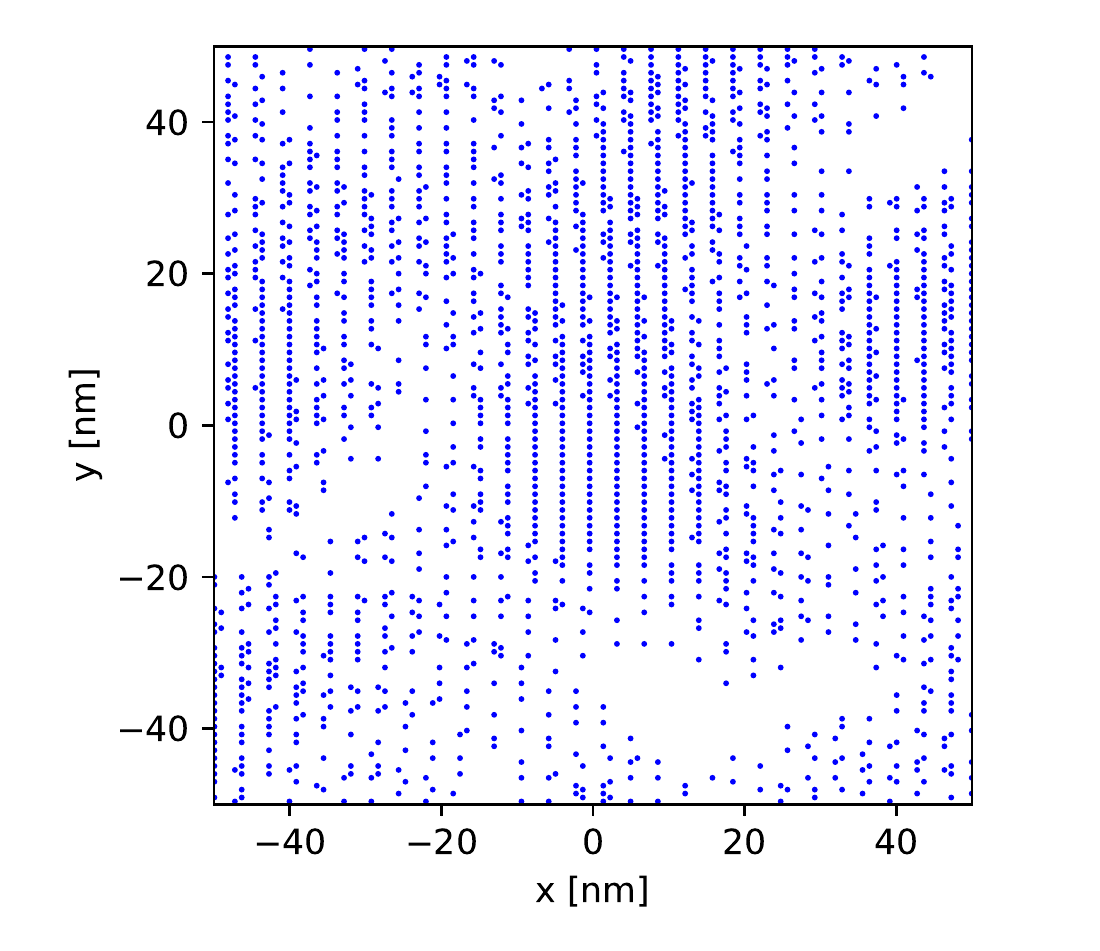}
    \caption{Mask cutout}
  \end{subfigure}
  \caption{\label{fig:kirchhoff_hex1} (a) Natural logarithm of the normalized intensity in the target plane obtained by simulations for a hexagonal mask illuminated by a realistic source. The mask has a distance between neighboring holes of \SI{0.9}{\nano\meter}, and the distance between the mask and target plane is \SI{40}{\micro\meter}. The desired target pattern is superimposed on the resulting target pattern, showing that the reproduction is true to size to within \SI{3}{\nano\meter}. (b) A $\num{100} \times \SI{100}{\nano\meter^2}$ cutout of the center of the mask.}
\end{figure*}

\begin{figure*}[htbp]
  \centering
  \begin{subfigure}[b]{0.4\textwidth} \includegraphics[width=\textwidth]{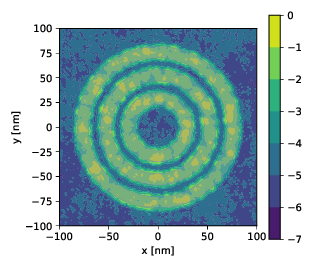}
    \caption{Intensity in target plane}
  \end{subfigure}
  \begin{subfigure}[b]{0.4\textwidth}  \includegraphics[width=\textwidth]{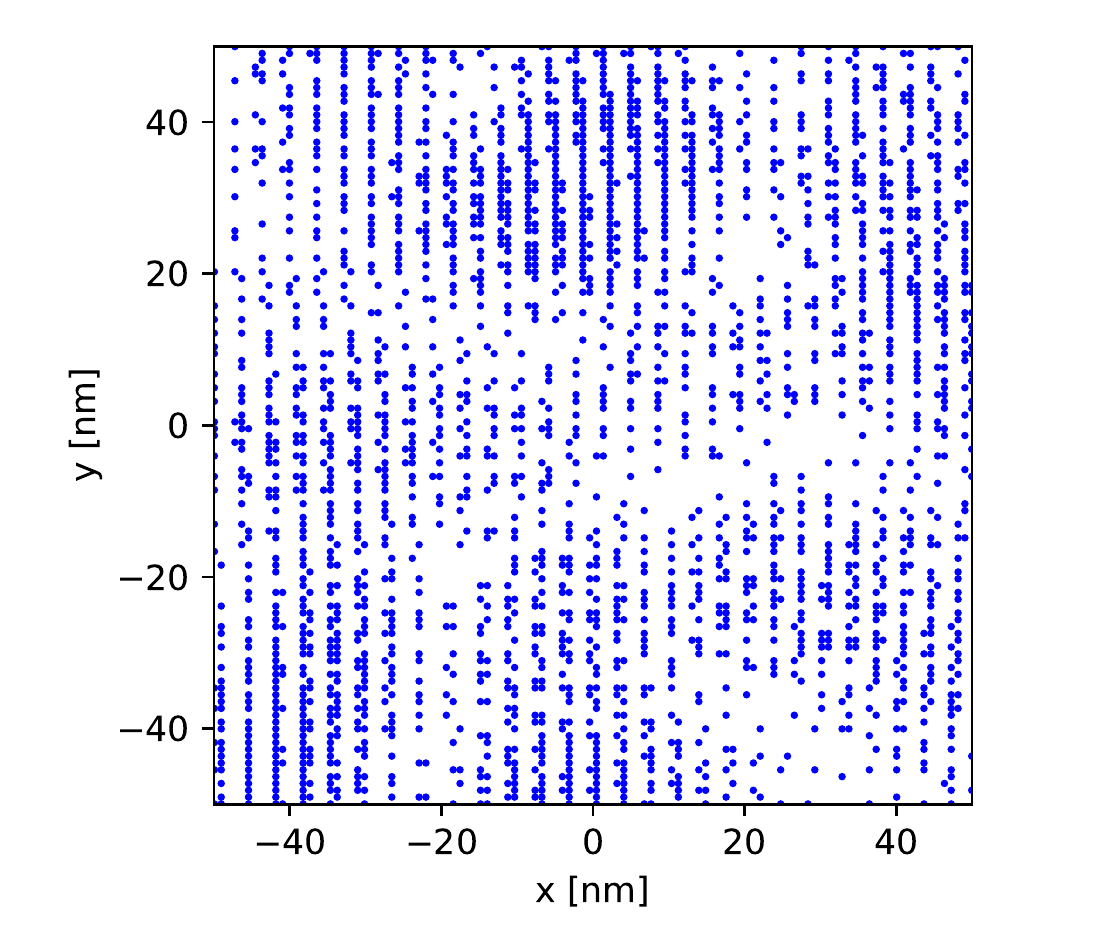}
    \caption{Mask cutout}
  \end{subfigure}
  \caption{\label{fig:kirchhoff_rect4} (a) Natural logarithm of the normalized intensity in the target plane obtained by simulations for a rectangular mask illuminated by a realistic source. The mask has a periodicity of \SI{0.9}{\nano\meter}, and the distance between the mask and target plane is \SI{40}{\micro\meter}. The intensity pattern shows a contrast of $0.83$. The desired target pattern is superimposed on the resulting target pattern. (b) A $\num{100}\times\SI{100}{\nano\meter^2}$ cutout of the center of the mask.}
\end{figure*}

To illustrate the versatility of the theoretical method outlined in the previous sections, we now present simulation results with different source and mask parameters. To this end, we use the two target test patterns presented in Fig.~\ref{fig:target}. The first pattern~[Fig.~\ref{fig:target}(a)] consists of a series of corner lines (or ``L shapes'') with width and spacing decreasing from \SI{10}{\nano\meter} to \SI{5}{\nano\meter}. This is a standard pattern used for resolution testing in lithography~\cite{Chang2001}. Most of the simulation results that we present are obtained under the assumption of this L-shaped pattern. It should, however, be noted that the L-shaped test pattern could, in principle, also be reproduced with a free standing mask by direct imaging. That is to say, for this particular pattern, holography is not strictly necessary. To illustrate the versatility of the near-field binary-holography method, we therefore, at the end of this section, include simulation results for a second test pattern, which consists of concentric circles~[Fig.~\ref{fig:target}(b)]. This pattern cannot be reproduced by direct imaging of a free-standing mask: the circular segments would fall out.

For all the simulations that we perform, the patterns are resolved into $512\times 512$ pixels and placed at a fixed distance $d=\SI{40}{\micro\meter}$ from the mask. For the beam we use an average wavelength of \SI{0.1}{\nano\meter}, which is typical for a helium beam as discussed in the Sec.~\ref{sec:introduction}. We use two different source configurations: (i) a plane wave (i.e., an ideal, monochromatic, perfectly coherent source) and (ii) a realistic source configuration as described in Sec.~\ref{sec:FK}, with source-mask distance of \SI{1.5284}{\meter} and skimmer diameter of \SI{400}{\micro\meter}. A wavelength distribution of $\lambda/\Delta\lambda=200$ has no appreciable impact on the results and is left out in the simulations presented here.

Masks can be made in two different ways: Holes can be ``drilled'' through a solid membrane as originally done by Fujita \etal~\cite{Fujita1996}. Alternatively, one can use a natural membrane and fill all of the undesired holes. This has the additional advantage of high precision because the position of the holes is built into the material. We chose our mask test parameters to reflect these two approaches. We chose one rectangular mask and one hexagonal mask, both with a hole-to-hole distance or periodicity of \SI{0.9}{\nano\meter}. This distance is chosen because it represents a natural limit of what one can imagine as possible for the density of holes in natural porous materials. The distance of \SI{0.9}{\nano\meter} is roughly the periodicity of beryl. Beryl is a silicate with a channel structure that allows individual atoms, typically metal and alkali ions, and water to be trapped in the channels. These trapped atoms were recently imaged for the first time using atomic resolution transmission electron microscopy~\cite{Ari2017}.
One may imagine that a mask could be made by deliberate filling of selected channels with atoms, with use of, for example, a focused ion beam. Quartz has similar channels, with a thinner wall structure, which would allow an even smaller periodicity. However, the thin wall means that quartz tends to become amorphous when prepared for thin-membrane experiments. While the channels in beryl and quartz are so small that they trap larger atoms and water molecules, they are still large enough to allow smaller atoms, for example helium, to penetrate~\cite{Swets1961}. Holes on the order of \SI{1}{\nano\meter} have recently been fabricated with use of helium-ion-beam lithography, which justifies the rectangular mask~\cite{Emmrich2016}. Most self-organized porous membranes, including beryl, have a hexagonal structure, hence the hexagonal masks.

The masks are designed by the grid-based subdivision method, with a minimal open-hole configuration~\cite{Nesse2017-2} to save computation time. The same masks are used for plane-wave and realistic sources.

The rectangular mask is designed to have an array of $512\times 512$ cells, with $4\times 4$ subcells, corresponding to a maximum $2048\times 2048$ hole openings. The hexagonal mask is made on a grid with $718 \times 718$ subcells. The mask is made from a hexagonally sampled mask field with $4 \times 3$ subdivisions in a cell by the method described in the Appendix.

Figure~\ref{fig:kirchhoff_rect1} show the simulation results for the plane-wave source.
The desired target pattern is superimposed on the actual target pattern, to illustrate how well the pattern generation works. We see that the target pattern is reproduced true to size to a precision of around \SI{2}{\nano\meter}. The contrast is $0.92$.

In Fig.~\ref{fig:kirchhoff_rect3} we present simulation results for the realistic source with the desired target pattern superimposed as before.
The target pattern is reproduced with a precision of  around \SI{3}{\nano\meter} and the contrast is \num{0.82}.

Figure~\ref{fig:kirchhoff_hex1} shows the results obtained with the
L-shaped pattern~[Fig.~\ref{fig:target}(a)] with a realistic source and a hexagonal mask. As expected, the pattern is reproduced just as well as with the rectangular mask. The only difference is a slightly lower contrast of \num{0.77}, which we attribute to the hexagonal mask having a smaller number of subcells in total.

Finally, in Fig.~\ref{fig:kirchhoff_rect4} we present portion of the mask and the simulated intensity in the target plane obtained with a mask designed for the purpose of creating the concentric circular pattern in Fig.~\ref{fig:target}(a). In obtaining this result, we use the realistic source model and we discretize the target pattern onto a rectangular array of $512\times 512$ points (as was done for the L-shaped pattern). The contrast of the intensity pattern is \num{0.83}.

Even though we have applied the near-field binary-holography technique to the successful design of the two specific patterns in Fig.~\ref{fig:target}, it should be stressed that the technique is, in principle, capable of creating arbitrary patterns. We believe that the results presented in Figs.~\ref{fig:kirchhoff_rect1}--\ref{fig:kirchhoff_rect4} testify to the potential of the approach that we propose.


\section{Throughput Estimation}
\label{sec:Throughput}

A high wafer throughput is a necessary requirement for chip mass production. It is therefore important to consider what writing speed we can hope to achieve with a  mask-based atom-lithography device and compare it with what is possible with EUV lithography. We chose here for comparison the NXE:3100 EUV tool from ASML. We use the numbers publicly available from ASML's websites~\cite{ASML2014}.

It is not trivial to make a suitable comparison. We decide simply to compare the photon flux (number of photons per second) and the atom flux (number of atoms per second). It is important to emphasize that this does not take into consideration resist performance, which is a very crucial factor. At the moment very little work has been done on resist development for metastable atoms.

We first calculate the flux for the EUV source. The power is \SI{10}{\watt}.
The energy of one EUV photon is $\SI{91.84}{\electronvolt}=\SI{1.47e-17}{\joule}$. Thus we get a flux of $\num{7e17}\mathrm{photons}/\text{s}$. 

The atom flux can be calculated on the basis of the numbers in Ref.~\cite{Even2015}. For a beam with optimum, narrow velocity distribution, the experimentally measured flux from the source is \num{3e16} atoms for a \SI{20}{\micro\second} pulse. The metastable discharge works with an efficiency of \num{1e-4} giving \num{3e12} atoms per pulse or {\num{1.5e17}} atoms/s during the pulse.

In this perspective the atom source is comparable in efficiency to the EUV source (\num{1.5e17} atoms/s versus \num{7e17} photons/s. However, in  practice the EUV source is superior, since  the valve in the pulsed source can currently operate at only \SI{300}{\hertz} while the EUV source is continuous. Also, newer EUV instruments with up to 25 times more power than given here are currently under development (NXE:3400B). Still, these simple calculations illustrate the promising potential of the metastable-atom sources for lithography.


\section{Conclusions and Outlook}
\label{sec:Conclusions}
In this paper we show how nanometer-resolution mask-based lithography of arbitrary patterns can be performed with realistic masks and atom sources. In addition we extend the binary-holography method to hexagonal cells and subcells, which makes it easier to calculate masks using self-organized porous membranes as mask substrates.

\begin{acknowledgments}
The authors gratefully acknowledge support from the Research Council of Norway, Fripro Project 213453 and Forny Project 234159. The research of I.S. was supported in part by the Research Council of Norway Contract No. 216699 and the French National Research Agency~(ANR) under contract ANR-15-CHIN-0003.
\end{acknowledgments}


\appendix 
\section{\label{app:hexgrid}Binary holograms on hexagonal grids}
\begin{figure}[htbp]
  \centering
  \includegraphics[width=0.5\columnwidth]{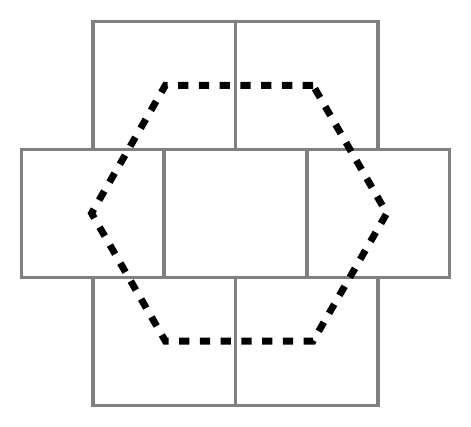}
  \caption{The Murphy and Gallagher approach: rectangular cells on a hexagonally sampled grid.}
  \label{fig:hex_mg_illustration}
\end{figure}

Murphy and Gallagher~\cite{Murphy1982} extended the binary-holography technique of Lohmann and Paris by placing the cells on a hexagonally sampled grid. The cells themselves are constructed as before, but their center points are placed on a hexagonal grid. This means that the cells are rectangular and that every other row of cells is shifted, something that must be taken into account when one is finding the correct location of each opening. The setup is illustrated in Fig.~\ref{fig:hex_mg_illustration}. The proposed use of hexagonally sampled fields would theoretically make more-accurate holograms because of the higher degree of symmetry available when circularly symmetric functions are represented. Computationally there are also potential advantages, first from a reduction in the amount of data stored and second from a reduction in the amount of work required to propagate the field between hexagonally sampled regions. Fourier methods of field propagation can be extended to work as efficiently on hexagonal grids as on rectangular grids with use of hexagonal fast Fourier transforms~\cite{Mersereau1979}.

\begin{figure}[htbp]
  \centering
  \includegraphics[width=\columnwidth]{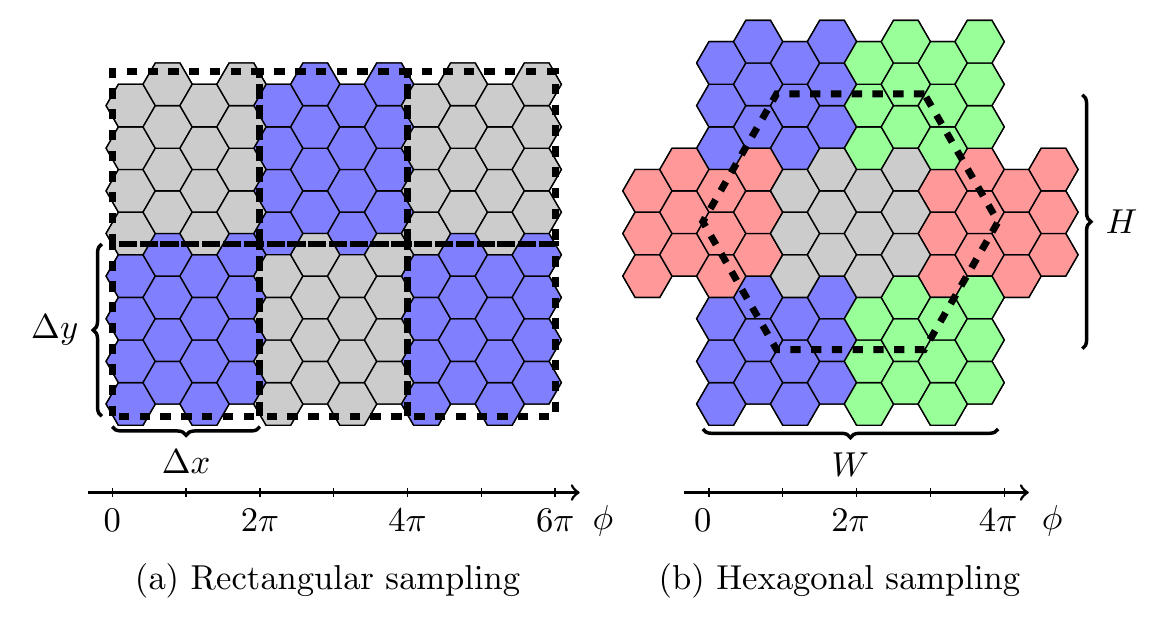}
  \caption{\label{fig:hex_illustration} New approach: hexagonal subcells on a hexagonal grid, with two possible sampling schemes (a) rectangular sampling and (b) hexagonal sampling.}
\end{figure}

Most self-assembled porous membranes have a hexagonal structure, and we are therefore concerned with adapting binary holography to work with a hexagonal lattice of holes. We discuss the adaptation of the grid-based holography method to work with subcells placed on a hexagonal grid. This can be done in a number of ways, but here we focus on two methods. The first method works by filling out a rectangular grid of cells using hexagonal subcells [Fig.~\ref{fig:hex_illustration}(a)], while the other method fills out a hexagonal grid of cells using hexagonal subcells [Fig.~\ref{fig:hex_illustration}(b)].

We first discuss performing grid-based holography with hexagonal subcells by starting from a rectangularly sampled grid. Let us consider a hexagonal grid with one edge of the hexagon parallel with the $x$ axis. The full width, $w$, of such a hexagon is related to its height, $h$, by
\begin{align}
  w &= \frac{2}{\sqrt{3}}h.
\end{align}
Every other column of the grid will be shifted along the $y$ axis a half height $h/2$. On such a grid we can create cells with $m\times n$ subcells to subdivide a rectangularly sampled grid with
\begin{align}
  \Delta x &= \frac{3}{4}wm,\\
  \Delta y &= hn.
\end{align}
In such a cell not all of the subcells will be aligned on the same rectangular grid, and if we assume a phase change along the $y$ axis when constructing the hologram this must be taken into account. If we have a phase change only along the $x$ axis, the contribution from each column will be the same. When such an approximation is performed, some of the subcells will overlap with the neighboring sample point. This is a natural extension of the ordinary method of grid-based holography, but it sets some limitations on the possible rectangular grids that can be used for sampling. An example of such a discretization scheme can be seen in Fig.~\ref{fig:hex_illustration}(a), with $m=4,~n=4$.

It is also possible to adapt the method of grid-based holography to hexagonally sampled mask fields, as illustrated in Fig.~\ref{fig:hex_illustration}(b). In the illustrated case we have also created rectangular cells of $m\times n$ subcells, but now the numbers $m$ and $n$ cannot be selected freely. The cells must have the correct proportions if they are to tile hexagonally, and be of the same shape and orientation. This is possible only for even numbers of $m$. If $m$ is odd, adjacent cells have to be flipped upside down, something that slightly moves their center. In Fig.~\ref{fig:hex_illustration}(b) we have $m=4$, $n=3$ which is one of the valid configurations. In this configuration, the repeating height is $H=2nh=6h$, and the repeating width is $W=(6/4)mw=6w$, and we see that both the width and the height are equally scaled.

Similarly to previous schemes, the next step is to assume a phase difference along the $x$ axis of $2\pi$ across every cell and then select in which columns to open subcells and how many subcells to open. The axis of phase change is shown below the graphics in Fig.~\ref{fig:hex_illustration}.

%
\nocite{apsrev41Control}
\bibliographystyle{apsrev4-1}
\bibliography{paper2018-03}

\end{document}